\documentclass[letterpaper, 10 pt, conference]{ieeeconf}  

\IEEEoverridecommandlockouts                              

\usepackage{graphics}       
\usepackage{graphicx}
\usepackage{epsfig}         
\usepackage{times}          
\usepackage{amsmath}        
\usepackage{amssymb}        
\usepackage{color}
\usepackage{array}
\usepackage{accents}

\usepackage{stackengine}
\usepackage{url}

\title{\LARGE \bf
Analysis of relay-based feedback compensation of Coulomb friction}

\author{Michael Ruderman, Leonid Fridman 
\thanks{This work has received funding from the EUs
H2020-MSCA-RISE research and innovation programme under grant
agreement No 734832. Second author acknowledges also support by
CONACyT (Consejo Nacional de Ciencia y Tecnologia), Project
282013; PAPIIT-UNAM (Programa de Apoyo a Proyectos de
Investigacion e Innovacion Tecnologica) IN106622
}%
\thanks{M. Ruderman is with University of Agder, 4979 Grimstad, Norway. %
        }%
\thanks{
L. Fridman is with Universidad Nacional Autonoma de Mexico,
Mexico.}%
\thanks{
\textcolor[rgb]{0.00,0.00,1.00}{\emph{Author's accepted
manuscript. To be published in IEEE VSS 2022.}}}
}

\begin{document}

\maketitle \thispagestyle{empty} \pagestyle{empty}

\begin{abstract}
Standard problem of one-degree-of-freedom mechanical systems with
Coulomb friction is revised for a relay-based feedback
stabilization. It is recalled that such a system with Coulomb
friction is asymptotically stabilizable via a relay-based output
feedback, as formerly shown in \cite{alvarez2000}. Assuming an
upper bounded Coulomb friction disturbance, a time-optimal gain of
the relay-based feedback control is found by minimizing the
derivative of the Lyapunov function proposed in \cite{sanchez2014}
for the twisting algorithm. Furthermore, changing from the
discontinuous Coulomb friction to a more physical
discontinuity-free one, which implies a transient presliding phase
at motion reversals, we analyze the residual steady-state
oscillations. This is in the sense of stable limit cycles, in
addition to chattering caused by the actuator dynamics. The
numerical examples and an experimental case study accompany the
provided analysis.
\end{abstract}

\bstctlcite{references:BSTcontrol}

\section{INTRODUCTION}
\label{sec:1}

The dry Coulomb friction in mechanical systems, whose resistive
force assumes either of two extreme values differing in the sign
opposite to the direction of relative motion, has always been a
'classical' but, at the same time, challenging example of
switching dynamics. This may also require a switching control, see
e.g. in \cite{utkin92}. The challenges in compensating for the
Coulomb friction are rooted in two facts. One is that it presents
a non-vanishing perturbation at zero equilibrium; and another
(even more challenging) one is that it is subject to eventual
discontinuities at the velocity zero crossing. Among the
model-free compensation methods aimed for attenuating the effect
of Coulomb friction on the controlled motion, the relay-based and
also sliding-mode-based strategies (see e.g. \cite{shtessel2014}
for basics) seem to be promising due to their robustness. Worth
mentioning that the characteristics of frictional systems are
mostly uncertain and/or perturbed, like for example the Coulomb
friction coefficient can be both time- and state-varying,
depending on such factors as the ambient temperature, surface
dust, normal load, dwell time, wear, lubrication and others.

In this work, we recall and extend the analysis of the relay-based
feedback compensation of the Coulomb friction provided in
\cite{alvarez2000}, however based on the same principles as the
widely celebrated sliding-mode twisting algorithm \cite{emel1986}.
Our goal is to analyze the convergence behavior of the stabilizing
relay-based compensator of the Coulomb friction and to derive the
time-optimal parametric conditions for its tuning. We are also
addressing the possible occurrence of the steady-state harmonic
oscillations and distinguish their causes by analyzing more
physical discontinuity-free friction transitions in the so-called
\emph{presliding} region of a motion initiation or motion
reversal. Therefore, both the rather classical discontinuous
Coulomb friction law and its tribology-justified,
discontinuity-free extension to a smooth force-displacement
mapping during presliding will be treated.

The rest of the paper is organized as follows. In section II, we
describe the considered class of mechanical systems with Coulomb
friction, thus formulating the problem statement. In section III,
we analyze the relay-based feedback control regarding its
time-optimal parametrization, while using analysis of the
continuous Lyapunov function provided in \cite{sanchez2014} for
the twisting algorithm. In section IV, we elaborate on the
residual steady-state oscillations, while introducing the
discontinuity-free Coulomb friction behavior and discussing the
appearance of the associated stable limit cycles. Also the
possible appearance of an inherent and, moreover, overlapped
chattering, in case of additional actuator dynamics, is recalled
in accord with \cite{boiko2004,aguilar2015}. In section V, we
present an illustrative experimental case study of compensating
the Coulomb friction. The theoretical and experimental results of
the paper are summarized and discussed in section VI.

\section{MECHANICAL SYSTEM WITH COULOMB FRICTION}
\label{sec:2}

We consider a feedback controlled one-degree-of-freedom mechanical
system with discontinuous Coulomb friction, i.e. $- C_f \,
\mathrm{sign}(x_2)$, which closed-loop dynamics is described by
\begin{equation}\label{eq:1}
\left(%
\begin{array}{c}
  \dot{x}_1 \\
  \dot{x}_2 \\
\end{array}%
\right) = \underset{\equiv g} {\underbrace{
\left(%
\begin{array}{cc}
0  & 1 \\
-k & -c -C_f/|x_2|  \\
\end{array}%
\right)
\left(%
\begin{array}{c}
  x_1 \\
  x_2 \\
\end{array}%
\right)
} }
 + \left(%
\begin{array}{c}
  0 \\
  1 \\
\end{array}%
\right) u.
\end{equation}
A state-feedback control, which is equivalent to a standard PD
(proportional-derivative) position controller, is parameterized
within the system matrix coefficients $k,c > 0$. A unity inertia
of the system is assumed for the sake of simplicity and without
loss of generality. The maximal (i.e. upper bounded) Coulomb
friction coefficient $C_f > 0$ is assumed to be known, and both
dynamic states of a relative motion $(x_1, x_2)(t)$ are assumed to
be available. Note that the input control channel $u$ is used for
the sake of friction compensation we are mainly interested in
here. Next, we will first show that an unforced system
\eqref{eq:1} will always approach the $x_1$-axis for all $t >
t_0$, independently of the initial conditions $t_0\geq 0$, $\|
x_1(t_0), x_2(t_0) \| < \infty $. When doing this, we will closely
follow the developments provided in \cite{alvarez2000}.

Assuming a positive definite and radially unbounded Lyapunov
function candidate $V_g(x_1, x_2) = 0.5 \, k x_1^2 + 0.5 \, x_2
^2$, and taking its time derivative, one obtains
\begin{equation}\label{eq:2}
\dot{V}_g = -c x_2^2 - C_f |x_2| < 0 \quad \forall \; x_2 \neq 0.
\end{equation}
Since $\dot{V}_g$ is negative definite everywhere except
$x_1$-axis, the unforced trajectories of the system \eqref{eq:1}
prove to reach always the manifold $\Lambda = \{(x_1,x_2) \in
\mathbb{R}^2: x_2 = 0 \}$ as $t \rightarrow \infty$. Note that
this is independent of the assigned control parameters $k > 0$ and
$c \geq 0$. Here $c > 0$ is in the sense of an exponentially
stable linear feedback system if $C_f = 0$; and $c = 0$ implies a
finite-time stability for the nonlinear damping $C_f > 0$. Next,
let us analyze the behavior of state trajectories when approaching
the $\Lambda$-manifold from two disjoint regions
$$
P^+ = \{x \in \mathbb{R}^2: x_2 > 0 \} \; \hbox{ and } \;  P^- =
\{x \in \mathbb{R}^2: x_2 < 0 \}
$$
of the $(x_1,x_2) \equiv x \in \mathbb{R}^2$ phase-plane. Since
the signum operator, which determines the Coulomb friction
dynamics, is defined in zero as $\mathrm{sign}(0) \in [-1, +1]$
(i.e. in the Filippov sense \cite{filippov1988}) the vector field
$g$ on the discontinuity manifold is
\begin{eqnarray}
\label{eq:3}
      g^{+} (x_{\Lambda}) & = {\underset{x \rightarrow x_{\Lambda},\: x \in P^{+}} \lim} g(x) = \left(%
\begin{array}{c}
  0 \\
  -k x_{1} - C_{f}
\end{array}%
\right), \\
\label{eq:4}
      g^{-} (x_{\Lambda}) & = {\underset{x \rightarrow x_{\Lambda},\: x \in P^{-}} \lim} g(x) = \left(%
\begin{array}{c}
  0 \\
  -k x_{1} + C_{f}
\end{array}%
\right).
\end{eqnarray}
From \eqref{eq:3}, \eqref{eq:4} it is visible that the velocity
vectors are pointing in opposite directions for $|x_{1}| \leq
C_{f} k^{-1}$. Since both vector fields are normal to the manifold
$\Lambda$, neither continuous motion nor sliding mode will occur
within
$$
\Lambda_I = \{(x_1,x_2) \in \mathbb{R}^2: \: -C_{f} k^{-1} \leq
x_{1} \leq C_{f} k^{-1}, \, x_2 = 0 \}.
$$
This forms the largest invariant set $\Lambda_I$ on the $x_1$-axis
which coincides with the range of residual control errors for the
class of motion systems \eqref{eq:1} perturbed by the Coulomb
friction with discontinuity. On the contrary, both vector fields
\eqref{eq:3}, \eqref{eq:4} are pointing in the same direction,
towards $P^{+}$ for $x_{1}< -C_{f} k^{-1}$ and towards $P^{-}$ for
$x_{1} > C_{f} k^{-1}$, in which way a continuous motion resumes
to take place.

The above results, cf. with \cite{alvarez2000}, are well known in
the control engineering practice, when an output feedback
controller is confronted with the issues of non-compensated
nonlinear friction. Note that adding an integral control part to
\eqref{eq:1} will not resolve convergence to stable zero
equilibrium, as has been recently demonstrated and discussed in
\cite{ruderman2021}.

\section{RELAY-BASED FEEDBACK}
\label{sec:3}

A stabilizing relay-based feedback control
\begin{equation}\label{eq:3:1}
u = -\gamma \, \mathrm{sign}(x_1)
\end{equation}
was analyzed in \cite{alvarez2000} for the second-order systems
with discontinuous Coulomb friction, cf. section \ref{sec:2}. Note
that in absence of the linear sub-dynamics in \eqref{eq:1}, i.e.
if $k,c=0$, the control system \eqref{eq:1}, \eqref{eq:3:1} will
coincide with the well-known twisting algorithm \cite{emel1986},
except that the velocity feedback part $-C_f \,
\mathrm{sign}(x_2)$ is no longer a control design term, but the
given inherent feature of the plant, namely the Coulomb friction,
cf. \eqref{eq:1}. It is well known that for an unperturbed double
integrator system, the twisting algorithm requires $\gamma > C_f$
for ensuring a finite-time exact convergence of
$(x_1,x_2)(t)$-trajectories to the stable zero equilibrium. When
the double integrator with relays feedback is augmented by the
residual system dynamics, i.e. $k,c \neq 0$, one still needs
analyzing the $\gamma / C_f > 1$ ratio for optimizing the
convergence time of the feedback compensator \eqref{eq:3:1},
provided the stable gains $k$ and $c$ are given. Below, we will
follow and use, while adapting for \eqref{eq:1}, \eqref{eq:3:1},
the Lyapunov function analysis of the twisting algorithm developed
and presented in \cite{sanchez2014}.

Introducing the upper and lower bounds $U > U^* > 0$ of the
matched control and perturbation quantities of \eqref{eq:1},
\eqref{eq:3:1} as
$$
U = \gamma + C_f - F, \quad U^*  = \gamma - C_f + F
$$
cf. \cite[section~III]{sanchez2014}, the continuous Lyapunov
function \cite{sanchez2014}
\begin{equation}\label{eq:3:2}
V(x) = \left\{%
\begin{array}{ll}
    \alpha_1 |x_2| + \alpha_2 \sqrt{x_2^2 + 2U |x_1|}, &  x_1 x_2 > 0 \\[1mm]
    \alpha_3 |x_2| + \alpha_4 \sqrt{x_2^2 + 2 U^*  |x_1|}, &  x_1 x_2 \leq 0. \\
\end{array}%
\right.
\end{equation}
can be used. Here, the collected coefficients are
$$
\alpha_1 = \frac{1}{U}, \: \alpha_2 = \frac{r^{-1}+r}{U(1-r)}, \:
\alpha_3 = -\frac{1}{U^*}, \: \alpha_4 = \alpha_1+\alpha_2 -
\alpha_3
$$
with $r= \sqrt{U^{-1} U^*}$. Beyond that, it was proved
\cite[Theorem~3.1]{sanchez2014} that the Lyapunov function's time
derivative is
\begin{equation}\label{eq:3:3}
\dot{V} \leq \left\{%
\begin{array}{ll}
    -1, &  x_1 x_2 > 0, \\[1mm]
   -\frac{\gamma - C_f - F}{\gamma + C_f + F}, &  x_1 x_2 < 0, \\
\end{array}%
\right.
\end{equation}
and the convergence time $T^*$ satisfies
\begin{equation}\label{eq:3:4}
T^* \leq \left\{%
\begin{array}{ll}
    V\bigl(x(0)\bigr), &  x_1(0) x_2(0) > 0, \\[1mm]
    r^2 V\bigl(x(0)\bigr), &  x_1(0) x_2(0) < 0. \\
\end{array}%
\right.
\end{equation}
It can be seen from \eqref{eq:3:3} that for the closed-loop system
to be asymptotically stable, the relay control gain must satisfy
$\gamma > C_f + F$. Furthermore, one can recognize from
\eqref{eq:3:4} that the convergence in the I-st and III-rd
quadrants of the phase-plane is not faster than in the II-nd and
IV-th quadrants, due to $r < 1$. We will make good use of this
fact when next analyzing the time-optimal feedback gain of a
relay-based Coulomb friction compensation. Further it is essential
to notice that the linear sub-dynamics in \eqref{eq:1} appears
instead of the $F$-bounded perturbation term of a double
integrator with feedback of both relays. Therefore, and without
lose of generality for \eqref{eq:1}, \eqref{eq:3:1}, we will set
$F=0$ and solely requires that the roots of the characteristic
equation $\lambda^2 + c \lambda + k = 0$ are all negative, where
$\lambda$ is the complex Laplace variable.

Since the convergence time estimates \eqref{eq:3:4} differ by the
multiplicative factor $r^2$, it is reasonable to look into the
convergence of those trajectories which start, correspondingly,
initially proceed in either II-nd or IV-th quadrant. Note that
from an application-related perspective it is fully in line with
the situation where an $x_1$-set-point is the control objective,
while $x_2(0)=0$ means there is no relative motion at the initial
time $t=0$. Evaluating the second case of \eqref{eq:3:2}, with
$F=0,\:x_2(0)=0$, and multiplying with $r^2$, cf. \eqref{eq:3:4},
one obtains the upper bound of the time estimate as
\begin{equation}\label{eq:3:5}
T \equiv \frac{\gamma \Bigl( 1 + \sqrt{\frac{\gamma - C_f}{\gamma
+ C_f}} \Bigr) }{\gamma C_f + C_f^2} \, \sqrt{2(\gamma-C_f)|x_1|}.
\end{equation}
Taking the derivative of \eqref{eq:3:5} with respect to $\gamma$
and evaluating numerically $\partial T / \partial \gamma = 0$, one
can find minima of the upper bound $T$ and, thus, the time-optimal
ratio $\gamma / C_f = 1.119$.

\begin{figure}[!h]
\centering
\includegraphics[width=0.88\columnwidth]{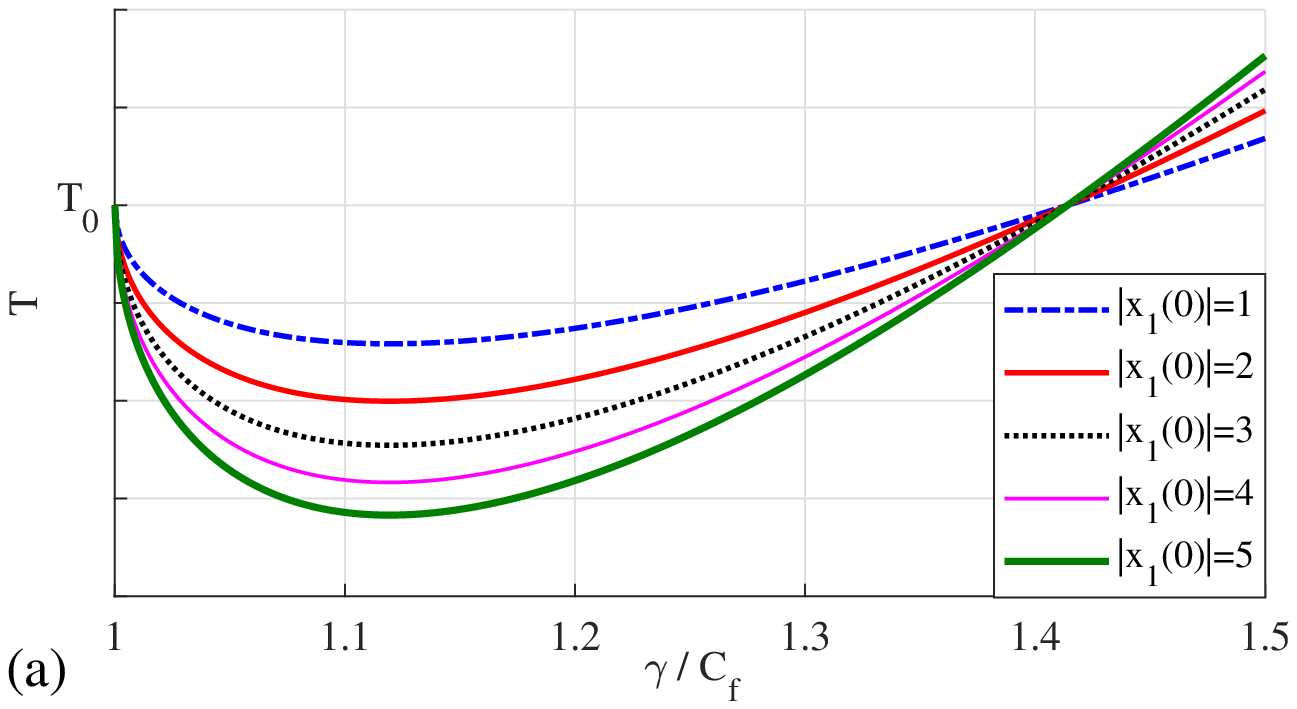}
\includegraphics[width=0.88\columnwidth]{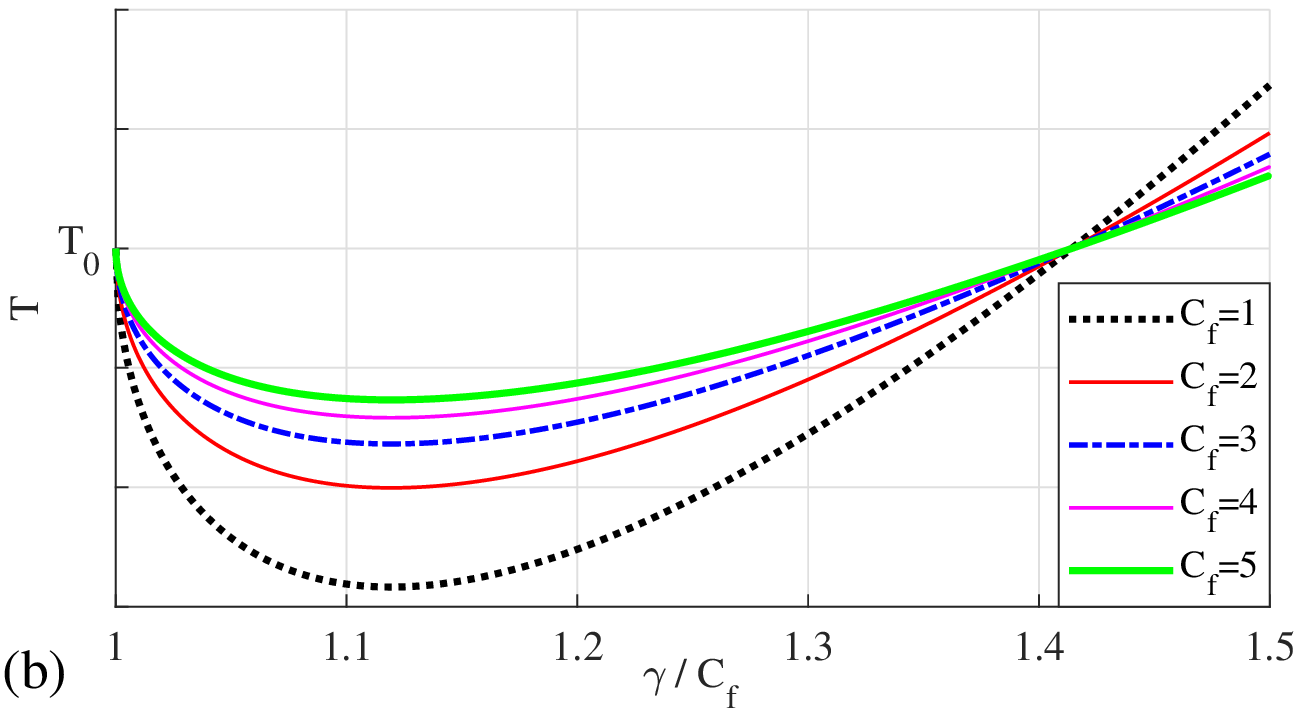}
\caption{The convergence time (unitless) upper bound versus the
$\gamma/C_f$ ratio: for the various initial conditions in (a), and
various $C_f$ values in (b).} \label{fig:3:1}
\end{figure}

The convergence time upper bound versus the control parameter
ratio $\gamma/C_f$ is shown exemplary in Fig. \ref{fig:3:1}, for
the different $x_1(0)$ initial values in (a) and Coulomb friction
coefficients $C_f$ in (b). Here $T_0$ denotes a boundary case
where the control gain $\gamma^+ \rightarrow C_f$ approaches
Coulomb friction coefficient from the right. One can recognize
that for selecting the time-optimal $\gamma$-gain, a possible
variation of the Coulomb friction parameter about the range of
40\% will not worsen the convergence performance comparing to
$T_0$.

\section{CONVERGENCE WITH LIMIT CYCLES}
\label{sec:4}

Next, we are addressing the problem of residual steady-state
oscillations which appear in form of the stable limit cycles. It
is worth noting that these should not be confused immediately with
the well-known \emph{chattering} effect in sliding-mode
applications, see e.g.
\cite{shtessel1996,bartolini1998,pisano2008}. The latter can well
occur upon the convergence of the control system (e.g. in twisting
mode) to zero equilibrium for a relative degree of the overall
plant (i.e. process and actuator) higher than two, see
\cite{boiko2004}. For ensuring that this is not a case for the
system \eqref{eq:1}, \eqref{eq:3:1} we will first briefly recall
the concluding statements of analysis provided in \cite{boiko2004}
for the second-order twisting algorithm. Then, we will introduce
the discontinuity-free transients of Coulomb friction, known as
\emph{presliding friction}, and address the occurrence of the
stable limit cycles owing to smooth transitions of the friction
force.

\subsection{Assessment of chattering by harmonic balance}
\label{sec:4:sub:1}

The describing function (DF) analysis, developed for the twisting
algorithm in \cite{boiko2004}, can be applied directly to the
system \eqref{eq:1} with discontinuous Coulomb friction and
relay-based feedback control \eqref{eq:3:1}. A parallel feedback
of two relay operators, one with the gain factor $\gamma$ and one
with the Coulomb friction coefficient $C_f$, allows writing the
DF, denoted by $\Omega(\cdot)$, as a sum of both DFs, i.e.
\begin{equation}\label{eq:40:1}
\Omega = \Omega_{\gamma} + j \omega \, \Omega_{C_f} = \frac{4
\gamma}{\pi A_1} + j \omega \frac{4 C_f}{\pi A_2} = \frac{4}{\pi
A_1} (\gamma + j C_f),
\end{equation}
where $\omega$ and $A_1$ are the angular frequency and amplitude
of the first harmonic of periodic oscillations of $x_1(t)$ at
steady-state, cf. \cite{boiko2004}. For convenience of the reader,
we recall that: (i) $A_2 = A_1 \omega$, due to the relationship
between $x_1$ and $x_2 = d x_1/dt$ in Laplace domain, and (ii) the
DF of an $\sigma$-amplified relay is $\Omega(A) = 4\sigma \, (\pi
A)^{-1}$. For proving the existence, correspondingly finding the
parameters, of the harmonic oscillations (i.e. chattering), the
corresponding harmonic balance equation
\begin{equation}\label{eq:40:2}
-\frac{1}{\Omega(A_1)} = G(j\omega) \equiv \frac{1}{-\omega^2 + j
\omega c + k}
\end{equation}
has to be solved. $G(j\omega) = x_1(j\omega)/u(j\omega)$ is the
input-to-output transfer function of the linear part of the system
\eqref{eq:1}, i.e. when excluding the Coulomb friction term out
from the states equation \eqref{eq:1}. Note that the graphic of
$-1/\Omega(A_1)$ is a straight line, which is starting from the
origin and progressing in negative direction (within II-nd
quadrant of the complex plane) as the amplitude $A_1$ increases.
The angle, correspondingly the slope, of that line is equal to
$\arctan(C_f/\gamma)$, cf. \cite{boiko2004}. Obviously, for any
physical (i.e. positive) parameter values of $k,\, c,\, \gamma,\,
C_f$, there is no intersection point of the Nyquist plot of
$G(j\omega)$ with $-1/\Omega(A_1)$ plot of the DF. That means no
$(\omega, A_1)$-solution of \eqref{eq:40:2} exists which implies
no harmonic oscillations can appear upon the convergence in
twisting mode of the control system \eqref{eq:1}, \eqref{eq:3:1}.

\subsection{Discontinuity-free Coulomb friction}
\label{sec:4:sub:2}

\begin{figure}[!h]
\centering
\includegraphics[width=0.75\columnwidth]{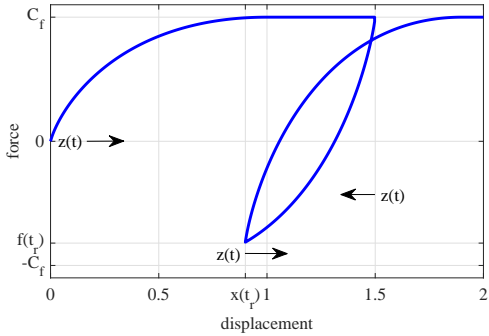}
\caption{Continuous transients of Coulomb friction at motion
reversals.} \label{fig:4:1}
\end{figure}

The discontinuity-free transients of the Coulomb friction at each
motion reversal (at time instant $t = t_r$) can be captured in
different manner, while fulfilling the rate-independency and
providing the so-called presliding hysteresis loops, cf. e.g.
\cite{ruderman2017}. Yet, according to the tribological study
\cite{koizumi1984}, the area of presliding hysteresis loops
increases proportionally to the 2nd power of the so-called
presliding distance $z(t)$, which captures the magnitude of
relative displacement after the sign of the relative velocity
changes, see \cite{ruderman2017} for details. Using the scaling
factor $s > 0$, which relates an after-reversal motion to the
presliding distance as
\begin{equation}\label{eq:4:1}
z = s \int \limits^{t}_{t_{r}} x_2 \, dt,
\end{equation}
defined on the interval $z \in [-1, \, 0) \cup (0, \, 1]$, one can
describe the branching of frictional force during presliding by
\begin{equation}\label{eq:4:2}
f_0(z)= z \bigl(1-\ln(z)\bigr).
\end{equation}
Each motion reversal at $t_r$ gives rise to a new presliding
transition captured by \eqref{eq:4:1}, \eqref{eq:4:2}, so that the
total (normalized) presliding friction map, cf.
\cite{ruderman2017}, is
\begin{equation}\label{eq:4:3}
f_p(t)= \bigl| \mathrm{sign}(x_2) - f_r \bigr| \, z
\bigl(1-\ln(z)\bigr) + f_r.
\end{equation}
The memory state of the last reversal transition is $f_r :=
f_p(t_r)$, cf. exemplified curves in Fig. \ref{fig:4:1}. Since the
presliding friction mapping \eqref{eq:4:3} is defined for $|z|
\leq 1$ only, the overall continuous Coulomb friction force is
given by
\begin{equation}\label{eq:4:4}
f(t)= \left\{%
\begin{array}{ll}
    C_f \, f_p(t) , & \hbox{if } |z|
\leq 1,  \\[1mm]
    C_f \, \mathrm{sign}\bigl(x_2(t)\bigr), & \hbox{otherwise.} \\
\end{array}%
\right.
\end{equation}

\subsection{Limit cycles in presliding range}
\label{sec:4:sub:3}

We now want to prove the occurrence of stable limit cycles around
zero equilibrium as a result of smooth frictional transitions upon
the motion reversals described above. For the sake of simplicity
and without loss of generality, we will in the following assume
$k, \, c=0$, so that the linear sub-dynamics is excluded as less
relevant for the main mechanisms of trajectory convergence and
emergence of the limit cycles. Following to that, the closed-loop
control system \eqref{eq:1}, \eqref{eq:3:1} is reduced to the
second-order dynamics
\begin{equation}\label{eq:4:4a}
\ddot{x}_1 + f(\dot{x}_1) + \gamma \, \mathrm{sign}(x_1) = 0.
\end{equation}
Assuming the Lyapunov function candidate $V_f = 0.5\, \dot{x}_1^2
+ \gamma \, |x_1|$, which is positive definite everywhere and
radially unbounded, one can easily obtain its time derivative as
\begin{equation}\label{eq:4:5}
\dot{V}_f = - \dot{x}_1 \, f(\dot{x}_1) .
\end{equation}

First, considering the discontinuous Coulomb friction, one can
directly show that
\begin{equation}\label{eq:4:6}
\dot{V}_f^d = - C_f \, | \dot{x}_1 | < 0 \quad \forall \quad
\dot{x}_1 \neq 0.
\end{equation}
This proves the system \eqref{eq:4:4a} will always reach
$x_1$-axis and is, thus, globally stable within invariant set
$\Lambda$, cf. section \ref{sec:2}. In order to demonstrate the
system \eqref{eq:4:4a} is but globally asymptotically stable, we
apply the LaSalle invariance principle and show that the invariant
set contains only one point, namely $\Lambda_e \equiv (0,0)$. All
trajectories starting from $\mathbb{R}^2 \setminus \Lambda_e$ will
reach $\Lambda$, at some time $t_1 > 0$, and then stay there for
all $t > t_1$, according to \eqref{eq:4:6}.
\begin{figure}[!h]
\centering
\includegraphics[width=0.85\columnwidth]{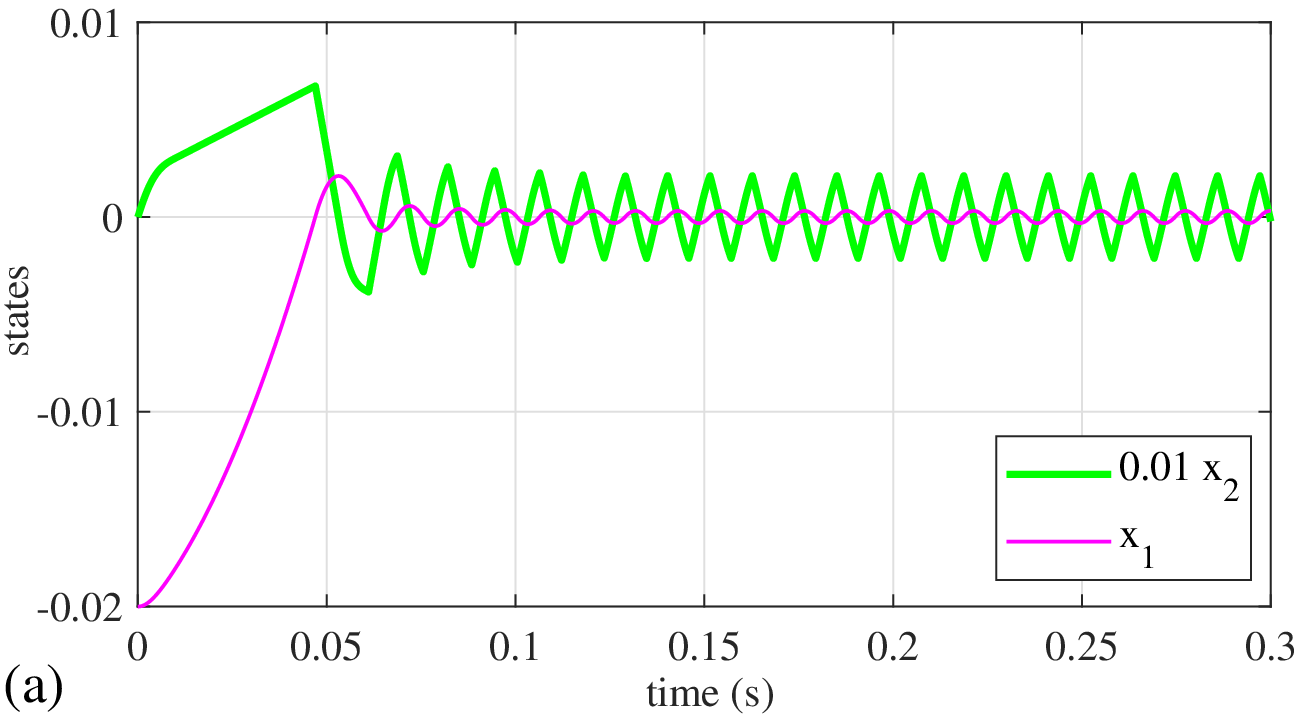}
\includegraphics[width=0.85\columnwidth]{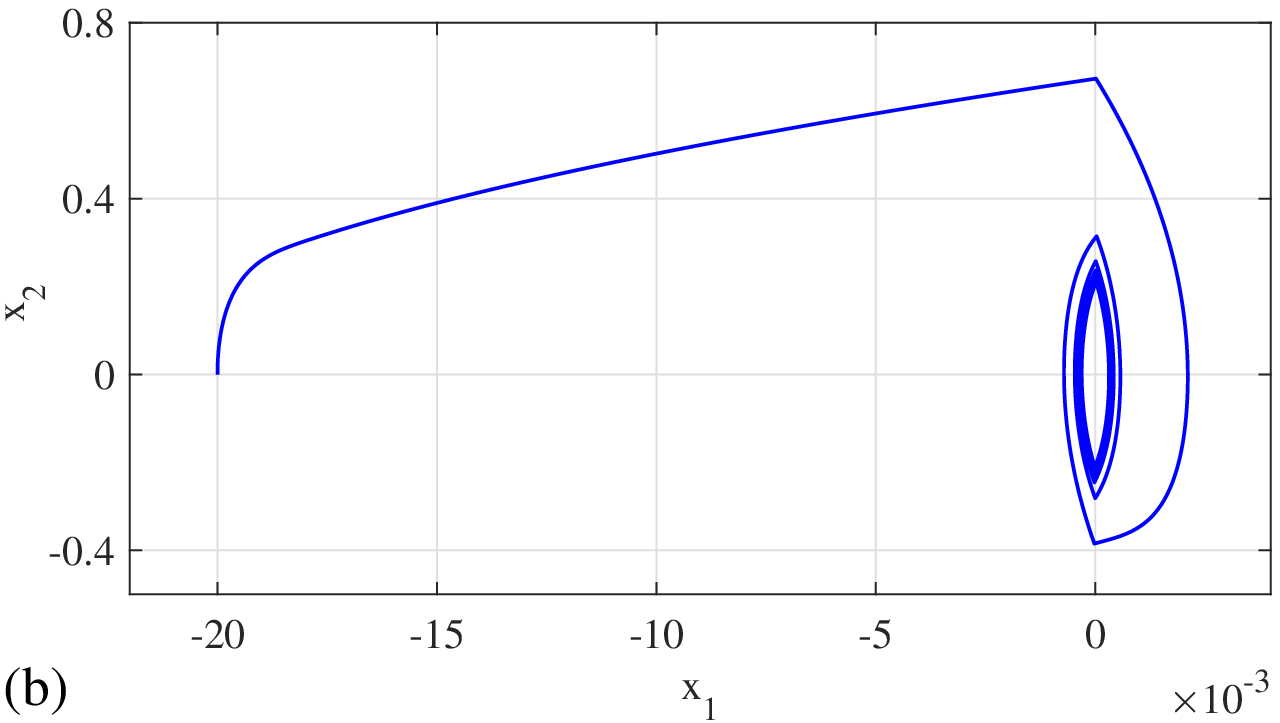}
\includegraphics[width=0.85\columnwidth]{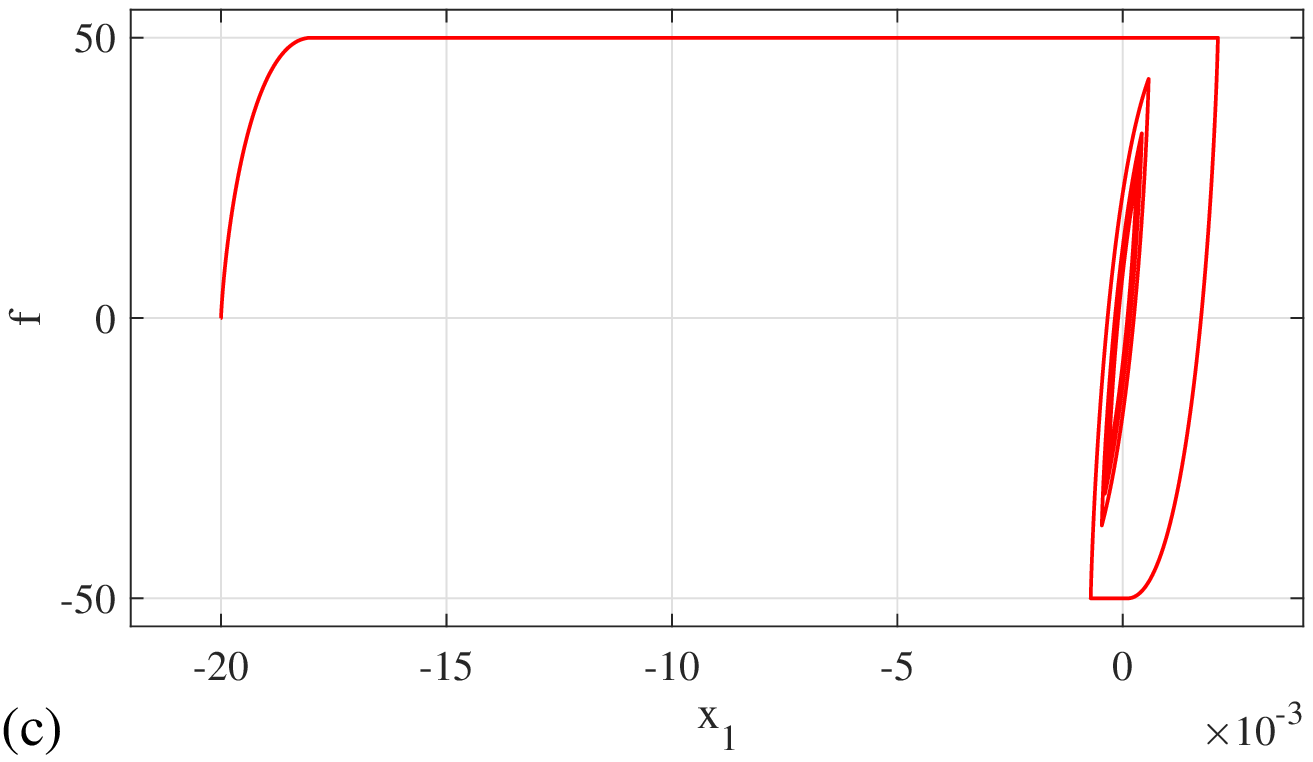}
\caption{Numerical example: position and velocity (down-scaled by
0.01) in (a), states trajectory in (b), and friction-displacement
curve in (c).} \label{fig:4:2}
\end{figure}
This implies $\ddot{x}_1 = 0 \; \Rightarrow \; C_f \,
\mathrm{sign}(\dot{x}_1) = - \gamma \, \mathrm{sign}(x_1)$. It is
evident that for all positive parameter values this equality is
impossible in the I-st and III-rd quadrant of the state-space, due
to the same signs. Thus, the above equality can hold only in the
II-nd and IV-th quadrant and \emph{iff} $\gamma = C_f$. However,
for any $\gamma > C_f$, cf. section \ref{sec:3}, the trajectory
will immediately move out of the set $\Lambda$, which is a
contradiction to its definition. Only in $\Lambda_e$, the
$\ddot{x}_1 = 0$ is fulfilled that proves the global asymptotic
convergence of the system \eqref{eq:4:4a} with $\gamma > C_f$ and
$f(\dot{x}_1) \equiv C_f \, \mathrm{sign}(\dot{x}_1)$. Note that
the Lyapunov function $V_f$ is weak, since not differentiable in
$x_1=0$. Thus, an extended invariance principle (see e.g.
\cite{utkin92}) must be applied for showing the asymptotic
stability. Indeed, one can show that the gradient $\partial V_f /
\partial x$ does not exist on the $x_1 = 0$ switching axis, which
implies $x_1 = \dot{x}_1 = 0$ is the unique solution, and there
are neither blocking nor unstable sliding modes on the manifold
$\bigl\{ x \in \mathbb{R}^2 \, | \, x_1 = 0 \bigr\} \setminus
\mathbf{0}$.

Next, assuming the continuous Coulomb friction \eqref{eq:4:4}, one
can show that in the presliding vicinity to zero equilibrium, the
time derivative is
\begin{equation}\label{eq:4:7}
\dot{V}_f^c = - C_f \dot{x}_1 f_p(z).
\end{equation}
Depending on the motion direction and the instantaneous sign of
the presliding transition curve $f_p(z)$, the time derivative of
the Lyapunov function candidate can be either positive or
negative, cf. \eqref{eq:4:7} and Fig. \ref{fig:4:1}. For
$\dot{V}_f^c (t) > 0$ the system will increase the energy level
and, in worst case, leave the presliding, meaning pass over to
$f(\dot{x}_1) = C_f \, \mathrm{sign}(\dot{x}_1)$. At the same
time, since the force-displacement transition curves $f_p(z)$ are
always clockwise, they behave as dissipative on each closed cycle.
That means the total $V_f^c(t_2) - V_f^c(t_1) = \Delta V_f^c  < 0$
for all $t_2 > t_1$ which are the two consecutive time instants of
a closed presliding cycle with $f_p(t_2) = f_p(t_1)$, cf. Fig.
\ref{fig:4:1}. Recall that the energy losses at such cycles are
equivalent to the corresponding area of the force-displacement
hysteresis loops. A stable limit cycle will occur once the input
energy (i.e by the control $u = - \gamma \, \mathrm{sign}(x_1)$)
supplied between two reversal time instants $t_2$ and $t_1$
becomes equal to $\Delta V_f^c$. The size of the limit cycle is $<
s^{-1}$ and depends on the $s$, $C_f$, and $\gamma$ parameters. An
exact analysis of the size and period of the limit cycles go
beyond the scope of the recent work and might be addressed in the
future research. An illustrative numerical example, with $C_f =
50$, $s=500$, $\gamma = 60$, and $k,c=0$, is shown in Fig.
\ref{fig:4:2}. Both system states are depicted as time series in
(a) and as a phase-plane trajectory in (b), while the friction
force over the relative displacement in (c), respectively. One can
recognize the appearance of a stable limit cycle, in accord with
the above discussion.

\section{EXPERIMENTAL CASE STUDY}
\label{sec:5}

\begin{figure}[!h]
\centering
\includegraphics[width=0.45\columnwidth]{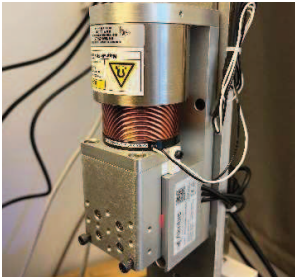}
\caption{Experimental setup of electro-mechanical drive
(laboratory view).} \label{fig:5:1}
\end{figure}

The relay-based feedback compensation of the Coulomb friction was
experimentally evaluated in the following case study, accomplished
on an electro-mechanical drive system in the laboratory setting,
cf. Fig. \ref{fig:5:1}.
The electro-magnetically actuated voice-coil motor drive has the
total linear stroke about 20 mm, which is indirectly measured by
the contactless inductive displacement sensor with a nominal
repeatability of $\pm 12$ $\mu m$. Note that due to a
position-varying amplification gain of the voice-coil motor and
the hardware-specific limitations of detection area of the
contactless sensor, a narrow displacement range of 6 mm only was
used in the following control experiments. The real-time board
operates the system with the set sampling rate of 10 kHz, while
the available control signal is the power-amplified voltage $V$ in
the range $[0,\,10]$ V. Further details of the setup can also be
found in \cite{ruderman2022}, with a main difference that the
additional oscillating payload is purposefully detached here from
the drive.

The constant gravitational term $mg$, where $m=0.735$ kg is the
overall moving mass and $g=9.8\, \mathrm{m}/\mathrm{sec}^2$ is the
constant of acceleration of gravity, is pre-compensated so that
the input-output system plant can be well described by the
second-order dynamics \eqref{eq:1}. Note that the nominal electric
time constant of 1.2 ms will be neglected.
\begin{figure}[!h]
\centering
\includegraphics[width=0.9\columnwidth]{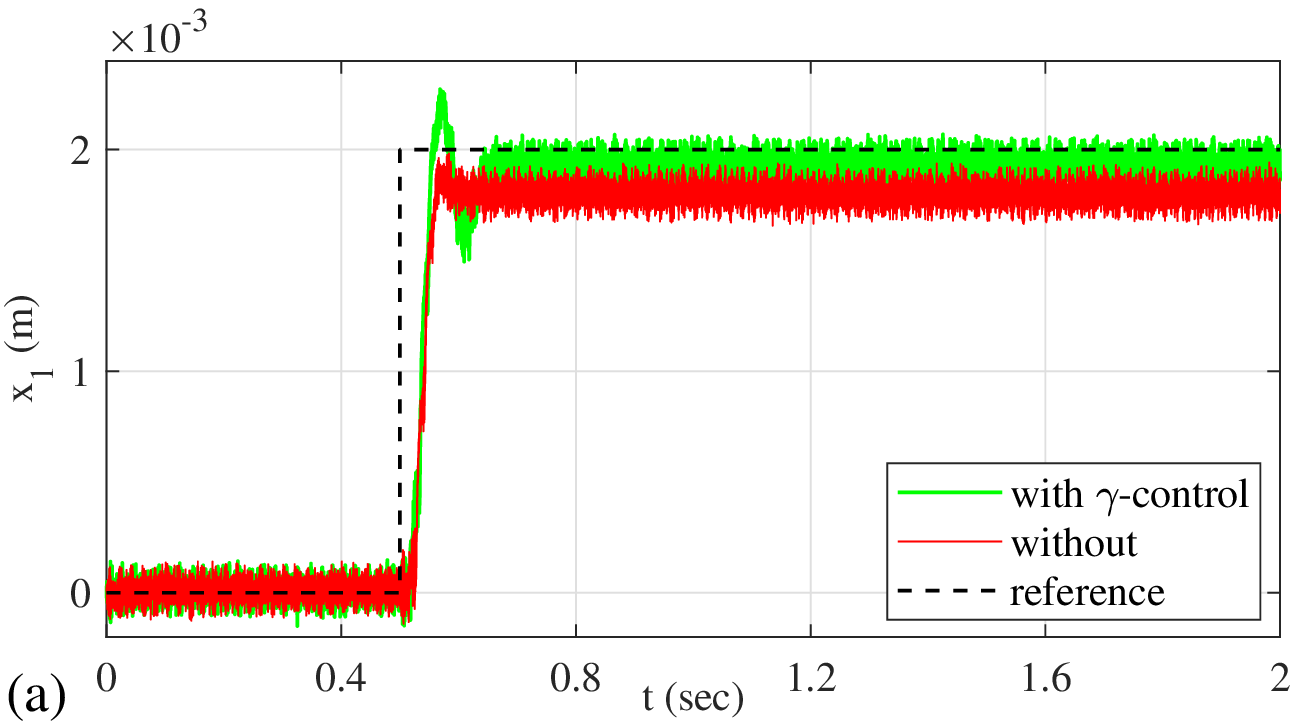}
\includegraphics[width=0.9\columnwidth]{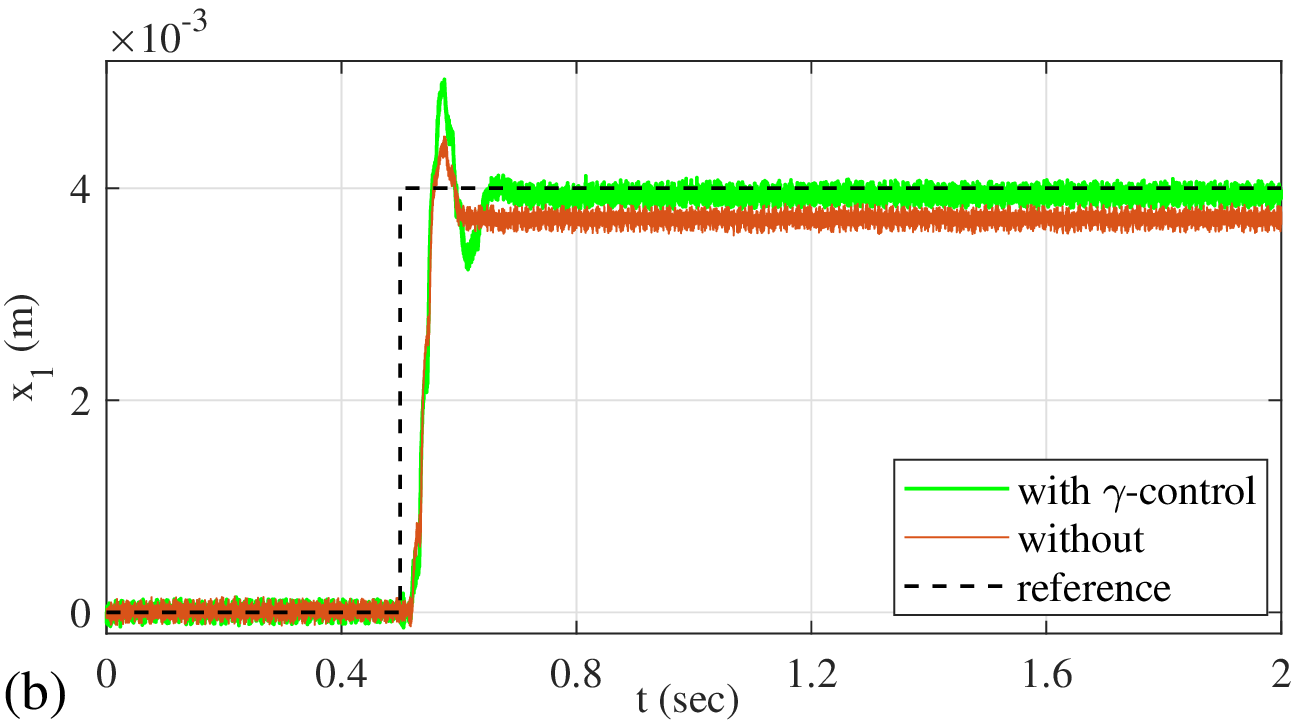}
\includegraphics[width=0.9\columnwidth]{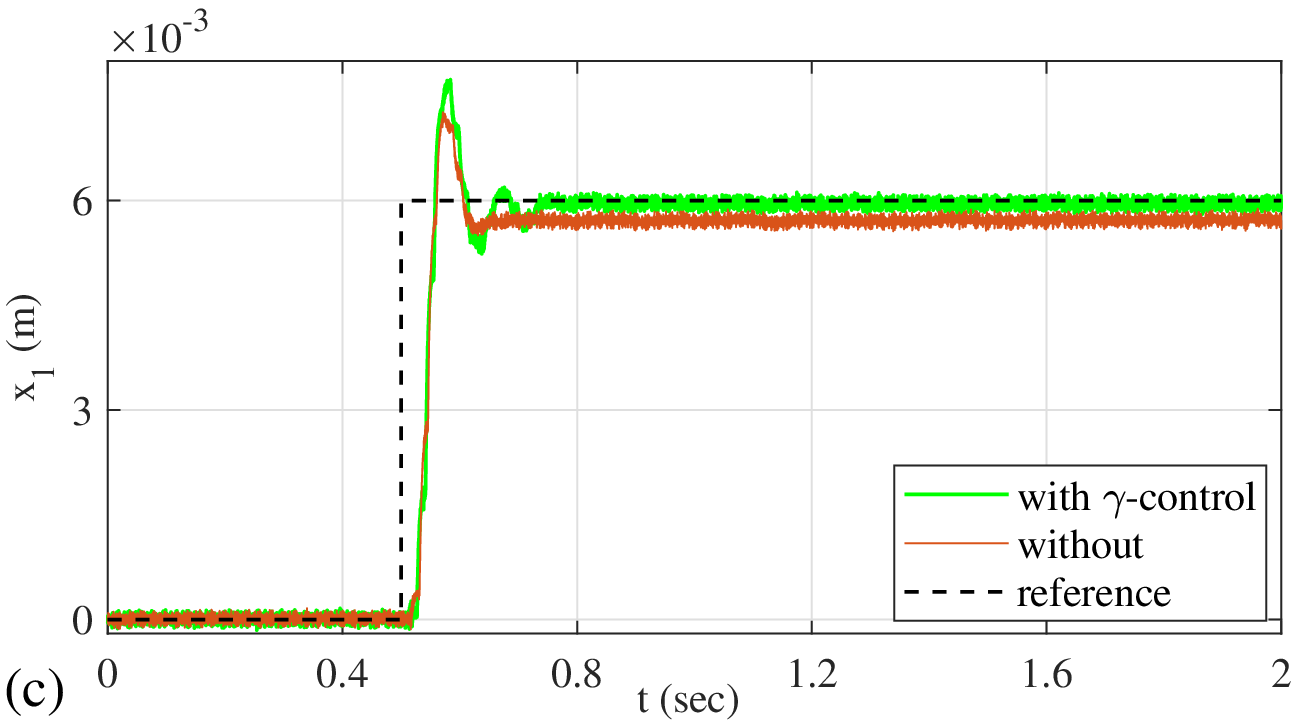}
\caption{Experimentally evaluated step response of the control
system with and without relay feedback compensator \eqref{eq:3:1},
with 2 mm reference step in (a), 4 mm reference step in (b), 6 mm
reference step in (c).} \label{fig:5:2}
\end{figure}
Therefore, we assume there is no additional actuator dynamics to
be taken explicitly into account, cf. with \cite{boiko2004}. The
system input signal is $u = \tau m^{-1} V$, where the motor force
constant $\tau=3.28$ is determined as the ratio between the
electromotive force constant and overall resistance of the coil
and connections. The identified Coulomb friction coefficient is
$C_f = 1.148$ N and the set linear feedback parameters are
$k=5600$ and $c=150$. The corresponding poles of the linear
sub-dynamics in \eqref{eq:1} are $\lambda_1 = -70$ and $\lambda_1
= -80$. It is worth noting that while the assigned $k$ is owned
entirely by the linear output feedback gain (respectively scaled
with $\tau m^{-1}$), the assigned $c$ value (equally scaled by
$\tau m^{-1}$) includes both, the identified viscous system
damping and the used output derivative feedback gain. The set
relay gain is $\gamma = 1.214$.

The experimentally evaluated step responses of the control system
with and without relay feedback compensator \eqref{eq:3:1} are
shown in Fig. \ref{fig:5:2}, for the step reference of 2 mm in
(a), 4 mm in (b), and 6 mm in (c). The $\gamma$-compensated
stiction in vicinity to the set reference, which is due to the
Coulomb friction, cf. section \ref{sec:2}, is visible in all three
experiments. For the time $t < 0.5$ sec, where no control action
appears, i.e. $V=0$, one can recognize a relatively high level of
the measurement noise. Note that a theoretically expected residual
position error range (when not compensating for friction cf.
section \ref{sec:2}) is about $\pm 0.2$ mm, according to the
$C_f/k$ ratio. The actually evaluated mean steady-state errors,
for $t > 0.8$ sec of all $\{2, \, 4, \, 6 \}$ mm references, are
$\{0.2, \, 0.29, \, 0.28 \}$ mm without $\gamma$-compensator, and
$\{0.07, \, 0.06, \, 0.04 \}$ mm with $\gamma$-compensator,
respectively. The residual control errors without
$\gamma$-compensator appear in line with the theoretical analysis,
while the residual control errors with $\gamma$-compensator are
close to the measurement noise and, moreover, due to additional
adhesion by-effects which are captured neither by discontinuous
nor discontinuity-free Coulomb friction laws.

\section{SUMMARY AND DISCUSSION}
\label{sec:6}

The relay-based feedback compensation of the Coulomb friction in
second-order systems with one mechanical degree of freedom was
analyzed. Since the closed-loop system with one output relay, due
to the compensator, and another output-rate relay, due to
discontinuous Coulomb friction law, is similar to the well-known
twisting-algorithm \cite{emel1986}, the existing approach of the
corresponding continuous Lyapunov function \cite{sanchez2014} was
used for the convergence analysis and determining a time-optimal
relay gain. Further, it was described why a more physical
discontinuity-free Coulomb friction, with the so-called presliding
transients, will unavoidably lead to the stable limit cycles in
vicinity to zero equilibrium. The uncompensated and feedback-relay
compensated regulation of the output displacement were
demonstrated in the experiments, where the residual steady-state
error (due to stiction) is in accord with the theoretically
expected one. Based on the results, reflections from analysis, and
interpretation of the numerical simulations and control
experiments, following points can be summarized for discussion.

\begin{itemize}

    \item[-] While the relative motion with discontinuous Coulomb friction is
    theoretically stabilizable by the output relay feedback,
    provided $\gamma > C_f$, a more physical actuator dynamics
    and, as a consequence, increase of the plant's relative degree to
    become $> 2$ will lead to harmonic oscillations of the output around equilibrium.
    This fact, which is in line with the DF-based analysis provided in
    \cite{boiko2004}, we observed experimentally when further increasing the
    $\gamma$-gain, even though it was admitted theoretically, cf. Fig.
    \ref{fig:3:1}. An increase of the $\gamma$-gain will reduce the
    inclination angle of the $-1/\Omega(A_1)$ slope, cf. section
    \ref{sec:4:sub:1}, and thus decrease the frequency and increase
    the amplitude of harmonic oscillations, as soon as the Nyquist plot
    of the overall plant with actuator will proceed also in the II-nd
    quadrant of the complex plane. Quantitatively it is directly visible
    since the amplitude of DF-determined harmonic oscillations is
    \begin{equation}\label{eq:6:1}
    A_1 = \frac{4}{\pi} \bigl| G(j \bar{\omega}) \bigr| \sqrt{\gamma^2 +
    C_f^2},
    \end{equation}
    cf. \cite{aguilar2015}, where $\bar{\omega}$ is the solution (if it exists)
    of the harmonic balance equation \eqref{eq:40:2}. Therefore, not only the
    knowledge of (otherwise generally uncertain) $C_f$ is relevant for an optimal tuning
    of the $\gamma$ control gain, but also the acceptance level
    of the residual output oscillations, taking into account the actuator dynamics.

    \item[-] Analysis of the discontinuity-free Coulomb friction behavior
    requires an accurate sensing of the relative displacement, with a
    possibly low level of both the measurement and process noise. On this account,
    the appearance of limit cycles due to presliding transitions could not be
    directly detected in the given experimental case study,
    despite such smooth force-displacement transitions are well known
    from the more accurate tribological investigations. An interesting point,
    which still requires both, more accurate position sensing
    and process knowledge, would be a decomposition of the residual output oscillations.
    Recall that it can include the components driven rather by additional
    actuator dynamics (i.e. due to relative degree $>2$) and those due to
    discontinuity-free Coulomb friction.

\end{itemize}

\bibliographystyle{IEEEtran}
\bibliography{references}

\end{document}